\newcommand{\be}{\begin{equation}}
\newcommand{\ee}{\end{equation}}
\newcommand{\ba}{\begin{eqnarray}}
\newcommand{\ea}{\end{eqnarray}}
\newcommand{\n}{\nonumber}
\newcommand{\p}{\partial}
\newcommand{{\ExponentialE }}{e}
\newcommand{\mathbb}[1]{\mbox{\bb #1}}

\newfont{\bb}{msbm10}
\newfont{\ff}{eufb10}

\newcommand{\op}[1]{\mbox{ $ \left(\frac{\partial}{\partial {#1}}\,\frac{1}{2 {#1}}
\,\frac{\partial}{\partial {#1}}\right)$ }}%

\documentclass[prl,superscriptaddress,12pt,tightenlines,preprint]{revtex4}

\usepackage{bm}
\usepackage{rotating}
\usepackage{graphicx}
\usepackage{epsfig}

\begin{document}

\title{Hydrogen atom in phase space: The Wigner representation}

\author{Ludmi{\l}a Praxmeyer}
\email{Ludmila.Praxmeyer@fuw.edu.pl} \affiliation{Instytut Fizyki
Teoretycznej, Uniwersytet Warszawski, Warszawa 00--681, Poland}

\author{Jan Mostowski}
\affiliation{Instytut Fizyki PAN, Warszawa 02--668, Lotnik\'ow 32/46 Poland}
\author{Krzysztof W\'odkiewicz}
\email{wodkiew@fuw.edu.pl} \affiliation{Department of Physics and
Astronomy, University of New Mexico, Albuquerque, NM~87131-1156,
USA} \affiliation{Instytut Fizyki Teoretycznej, Uniwersytet
Warszawski, Warszawa 00--681, Poland}

\date{\today}

\begin{abstract}
We have found an effective method of calculating the Wigner
function, being a quantum analogue of joint probability
distribution of position and momentum, for bound states of
nonrelativistic hydrogen atom. The formal similarity between the
eigenfunctions of nonrelativistic hydrogen atom in the momentum
representation and Klein-Gordon propagators has allowed the
calculation of  the Wigner function for an arbitrary bound state
of the hydrogen atom. These Wigner  functions for some low lying
states are depicted and discussed.
\end{abstract}


\maketitle

There are many ways to visualize quantum states of a single
particle. Wave functions in position representation are a
natural tool to present probability distribution in the
configuration space. On the other hand it is easier to
visualize momentum distribution if the momentum
representation of the wave function is used. Phase space
description, namely Wigner function of a state, provides a
natural generalization of joint position and momentum
distribution.  Wigner function corresponding to a state
described by wave function $\psi(\vec{r})$ is defined as
follows \cite{Wigner}:
\begin{eqnarray}
\label{wignerdef}  W_{\psi}(\vec{r},\vec{k})&=& \int\;
\frac{d^3 q}{(2\pi)^3} \,\psi^{\ast} (\vec{r}+\vec{q}/2) \,
e^{i\vec{q}\vec{k}}\, \psi( \vec{r}- \vec{q}/2)
 \nonumber\\
&=& \int\; \frac{d^3 q}{(2\pi)^6} \,\tilde{\psi}^{\ast}
(\vec{k}+\vec{q}/2)\, e^{-i\vec{q}\vec{x}}\, \tilde{\psi}(
\vec{k}- \vec{q}/2)\,
\end{eqnarray}
As it is well known the Wigner function for most quantum
states is non-positive and bounded to an interval
$\frac{1}{\pi^3}[ -1,1 ]$, with marginals in $\vec{r}$ and
$\vec{k}$ corresponding to momentum (wave-vector) and
position quantum probability distributions. In the
classical limit the Wigner function becomes a classical
phase space distribution. These and other properties of the
Wigner function with its applications in various branches
of physics have been reviewed in a number of articles and
books
\cite{wignerreview}.

It is well known that Wigner function given by Eq.
(\ref{wignerdef}) can be easily calculated, analytically or
numerically, for most of one dimensional systems. In case
of three dimensional problems, especially the ones with
spherical symmetry, calculations are usually much more
difficult. Integrals become quite cumbersome and in most
cases impossible to handle analytically. Despite the
existence of analytical expressions for the hydrogen atom
wave functions in position and momentum representations
\cite{hydrogen}, the form of the phase space Wigner
function is unknown. Analytical formula for the Wigner
function is not even known for the {\bf 1s} state of the
hydrogen atom. In the literature one can find only a
limited number of papers devoted to this subject
\cite{Dahl1,Dahl2}, and some of the published results have
been achieved using some kind of approximate methods that
have generated controversies
\cite{Nouri}. In a different context, the hydrogen atom has been
investigated recently using the Kirkwood-Rihaczek phase
space representation, which is easy because it involves
only products of the momentum and  position wave functions
with a proper phase \cite{wodRih}.

Due to the discovery of quantum phase space tomography,
Wigner functions have been experimentally reconstructed for
quantum states of light, vibrational modes of molecules,
and superpositions of diffracted cold atoms by a double
slit
\cite{tomoreview}. It has been discovered recently that the
phase space plots of the Wigner function provide a unique
visualization of the quantum state that can unravel such
unique quantum properties like entanglement of correlated
systems \cite{bgekw} or the phase sub-Planck structures of
quantum interference \cite{zurek}. Because of all these
reasons an analytical formula for the phase space Wigner
function, of such a fundamental system like the hydrogen
atom, can be useful for quantum tomography, quantum state
diagnostic and phase space visualization of negative
structures of quantum interference.

This rather vexing situation regarding the analytical form
of Wigner functions for the hydrogen atom indicates that
the calculation of the phase space representation for the
hydrogen requires a new approach based on a new method or a
calculational trick to overcome the old difficulties. It is
the purpose of this Letter to present an analytical
computational scheme for the calculation of the phase space
Wigner function for arbitrary bound-energy eigenfunction of
the hydrogen atoms in terms of a set of generating
differential operators acting on a simple single integral.
This general ``hydrogen atom integral'' (HAI), dependent on
the Bohr radius and other geometrical parameters of the
hydrogen eigenfunction, can be easily calculated
numerically.

Before we present the outline of the technical features of
the general approach, we first summarize our main results,
and illustrate the power of our methods showing for the
first time exact phase space plots of the hydrogen in the
Wigner representation. The main result of our paper can be
written in the form of the following formula for the phase
space Wigner function for the hydrogen energy eigenvectors:
\begin{equation}\label{genformula}
W_{\psi_{nlm}}(\vec{r},\vec{k}) =
\mathbb{D}_{nlm}\bigg(\frac{\partial}{\partial
\vec{k}},\frac{\partial}{\partial b_1},\frac{\partial}{\partial
b_2} \bigg)\; I(r,k,\vec{r}\vec{k},b_1,b_2)\bigg{|}_{b_1=b_2=1/na}\,.
\end{equation}
In this formula $\mathbb{D}_{nlm}$ is a linear differential
operator reproducing the Wigner function for {\it all}
hydrogen states from a simple hydrogen atom integral  (HAI)
defined below by Eq. (\ref{hai}). The HAI depends on three
scalars only: $r=|\vec{r}|$, $k=|\vec{k}|$ and
$\vec{r}\vec{k}=rk
\cos\theta$. The two arbitrary running parameters $b_1$ and
$b_2$, are determined at the end of the calculations only
by $1/na $, where $n$ is the principal quantum number and
$a$ is the Bohr radius. The form of this HAI is
\begin{equation}\label{hai}
I(r,k,\vec{r}\vec{k},b_1,b_2)=\int_0^1 du \exp\bigl(4i
u\vec{r}\vec{k}\bigr) \frac{1}{C(u)} \exp\bigl(-2 r
C(u)\bigr)\,
\end{equation}
where
\begin{equation}\label{Cdef}
C(u)=\sqrt{ub_{1}^{2}+(1-u)b_{2}^{2}+4u(1-u) k^2}\,.
\end{equation}
The formula is the central result of our paper. It can be
used to generate the Wigner function for an arbitrary
hydrogen energy eigenfunction. The HAI plays a role of a
generating function for all Wigner functions of the
hydrogen atom.  Differential operators $\mathbb{D}_{nlm} $
acting on HAI give Wigner functions for all bound states of
hydrogen atom.

Before we explain how this central result has been
obtained, we write down the formula for the Wigner function
of the ground state. For the $\bf{ 1s}$ state this operator
reads:
\begin{equation}\label{d100}
\mathbb{D}_{100}=\frac{2 e^{-2i\vec{k}\vec{r}}}{\pi^3 a^3}
\frac{\partial^2}{\partial (b_1^2)\partial (b_2^2)}
\end{equation}
and the corresponding Wigner function (\ref{genformula}) is
\begin{eqnarray}
W_{\psi_{100}}(\vec{r},\vec{k})=\frac{2
e^{-2i\vec{k}\vec{r}}}{\pi^3 a^3} \frac{\partial^2}{\partial
(b_1^2)\partial (b_2^2)}
I(r,k,\vec{r}\vec{k},b_1,b_2)\bigg{|}_{b_1=b_2=1/a}\,.
\label{w100}
\end{eqnarray}
We can perform all the derivatives and write the Wigner
function as a single integral, but this lengthy formula is
rather useless for this letter, and we omit it.
We note that the Wigner function for the ground state
depends only on three scalars $r,k$ and $\theta$. A very
simple numerical calculations of the HAI leads to the
Wigner function for the ground state.

\begin{center}
\begin{figure}[h]
\begin{picture}(0,0)(35,10)
\put(100,137){\makebox(0,0){$\theta=0 $}}
\put(220,135){\makebox(0,0){$\theta=\pi/4 $}}
\put(345,135){\makebox(0,0){$\theta=\pi/2 $}}
\end{picture}
\includegraphics[scale=.6]{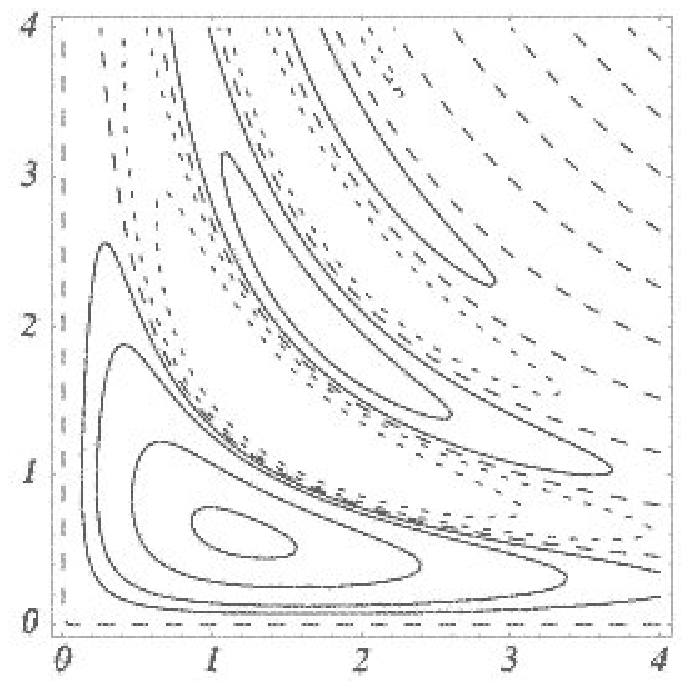} $\;$
\includegraphics[scale=.6]{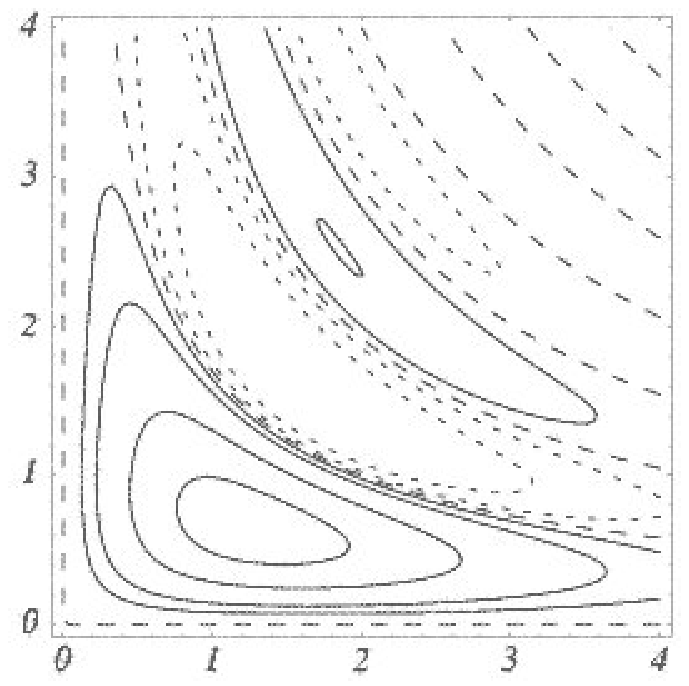} $\;$
\includegraphics[scale=.58]{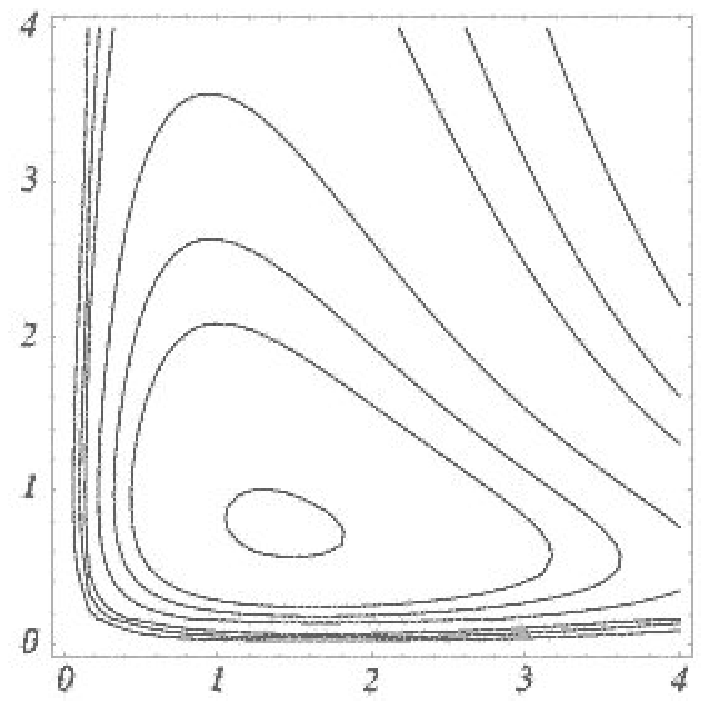}
\caption{\small  $1s$ state: $4\pi r^2 k^2
W_{1s}(\vec{r},\vec{k})$. Contour plots to be compared with
those from reference \cite{Dahl1}. Dashed lines denote a zero-level,
dotted lines denote negative values, separate distance between the
contours are chosen as in \cite{Dahl1}.
}\label{fig1}
\end{figure}
\end{center}

In Figure
(\ref{fig1}), we have depicted contours of $4\pi r^2 k^2
W_{100}(r,k,\theta)$ for selected values of $\theta$. These
figures should be compared with the only published
numerical results obtained 22 years ago in  reference
\cite{Dahl1}.
In the following Figure (\ref{fig2}) we have depicted the
same function $ W_{100}(r,k,\theta)$  multiplied by factor
$r^2k^2$ for $\theta=0$ and $\theta=\frac{\pi}{2}$.  We see
explicitly regions of the phase space where the Wigner
function is non-positive.

\begin{figure}[h]
\begin{picture}(0,0)(35,10)
\put(80,160){\makebox(0,0){$\theta=0 $}}
\put(300,160){\makebox(0,0){$\theta=\pi/2 $}}
\end{picture}
\includegraphics[scale=.8]{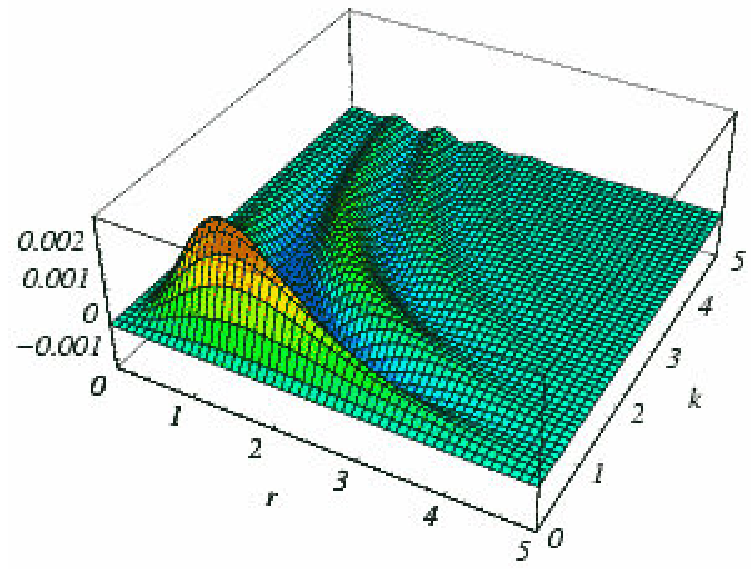}$\quad$
\includegraphics[scale=.8]{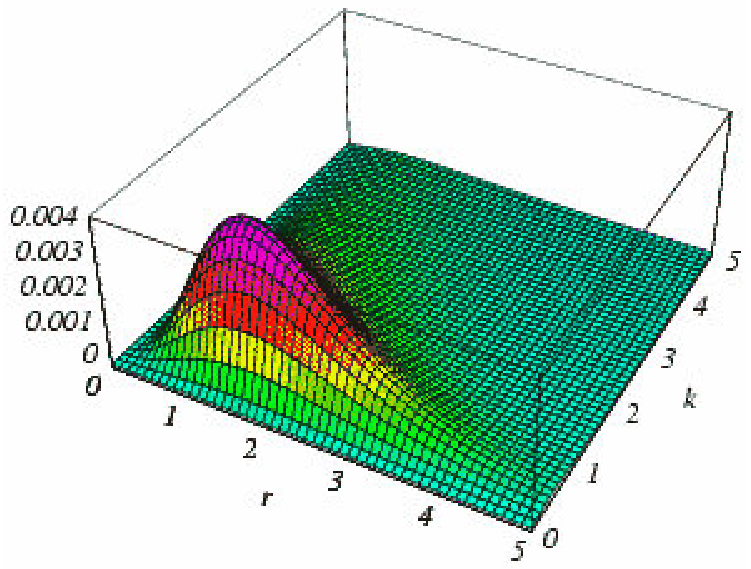}
\caption{\small The Wigner function of $1s$ state multiplied by factor
$r^2 k^2$. Only  the cross-section for
$\theta=\frac{\pi}{2}$ is a positive function, for all
others values of the angle between position and momentum
vectors cross-sections have negative values. For $\theta=0$
we see explicitly an oscillating structure.}\label{fig2}
\end{figure}

\begin{center}
\begin{figure}[ht]
\begin{picture}(0,0)(35,10)
\put(70,150){\makebox(0,0){$\theta=0 $}}
\put(270,150){\makebox(0,0){$\theta=\frac{\pi}{2} $}}
\put(30,-12){\makebox(0,0){$\theta=0 $}}
\put(212,-12){\makebox(0,0){$\theta=\frac{\pi}{4} $}}
\end{picture}
\includegraphics[scale=.9]{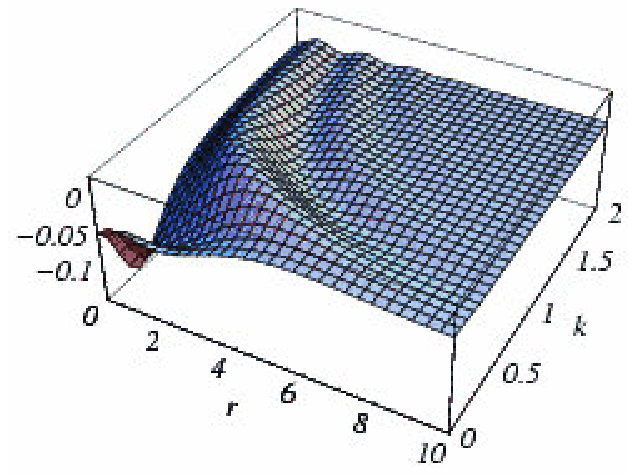}$\qquad$
\includegraphics[scale=.75]{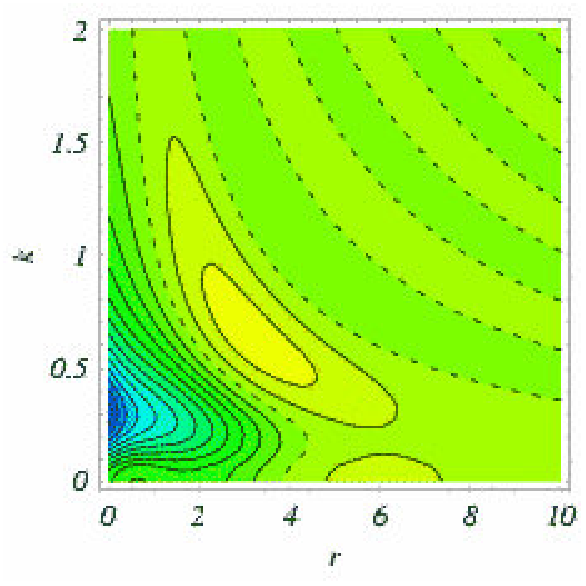}\\
$\;\;\;\;$\includegraphics[scale=.75]{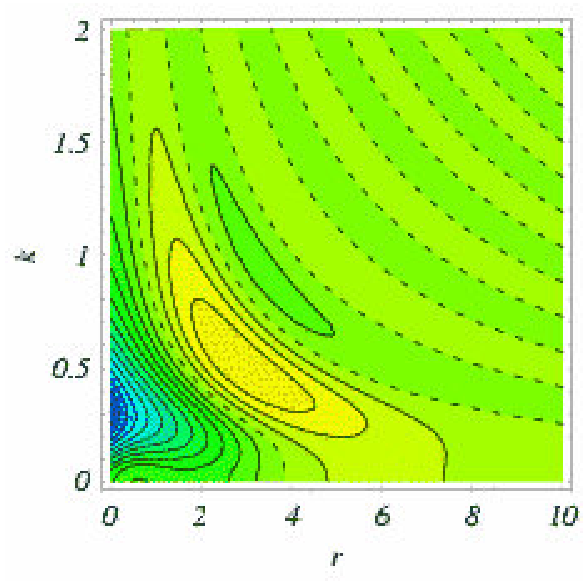}$\qquad\qquad\quad$
\includegraphics[scale=.75]{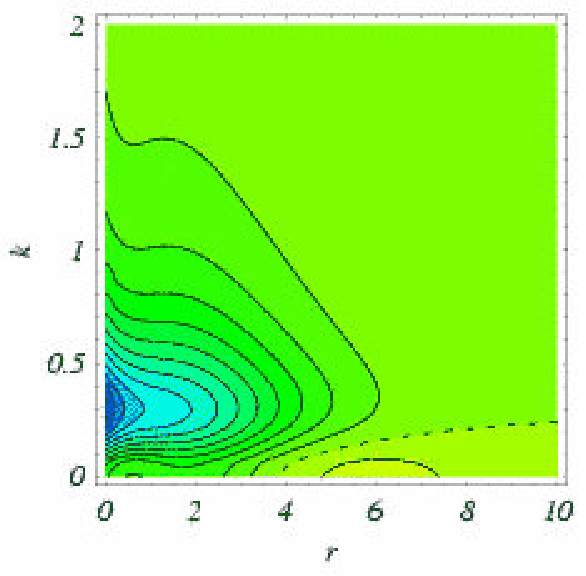}
\caption{\small The Wigner function of the $2s$ state. Dashed lines
denote a zero-level. Other contours are separated by $0.01$}
\label{fig3}
\end{figure}
\end{center}

The power of our method can be exhibited further if higher
states of the hydrogen atom are considered. Below we just
quote the corresponding differential operators for the
$\bf{ 2s}$ state:          
\begin{equation}\label{d200}
\mathbb{D}_{200} =
 -\frac{e^{-2i\vec{k}\vec{r}}}{4\pi^3 a^3}
\left[\frac{\partial}{\partial b_1}-b_1\frac{\partial^2}{\partial
b_1^2}\right]
\left[\frac{\partial}{\partial b_2}-b_2\frac{\partial^2}{\partial
b_2^2}\right]
\end{equation}
where after all the calculations we put $b_1=b_2 = 1/2a$.
With a little patience or help from a symbolic software all
the differentiations of the HAI can be performed, and a
close form expression for the Wigner function
$W_{\psi_{200}}(r,k,\theta)$  from Eq. (\ref{genformula})
can be calculated and plotted. In Figure (\ref{fig3}) we
present contour plots of the $2s$ Wigner function for
various values of $\theta$. Again, the negative values of the Wigner
function are clearly seen. 

%

\begin{center}
\begin{figure}[h]
\begin{picture}(0,0)(35,10)
\put(30,40){\makebox(0,0){ a)}}
\put(250,40){\makebox(0,0){b)}}
\end{picture}
\includegraphics[scale=.9]{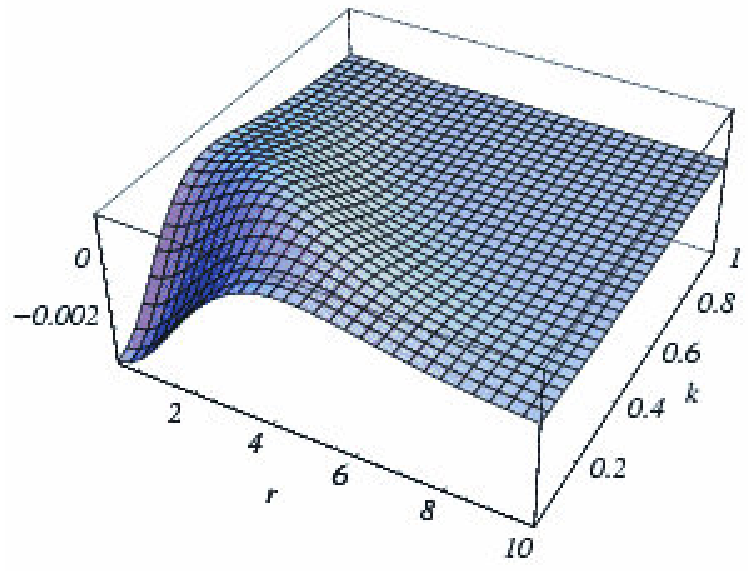}$\qquad$
\includegraphics[scale=.9]{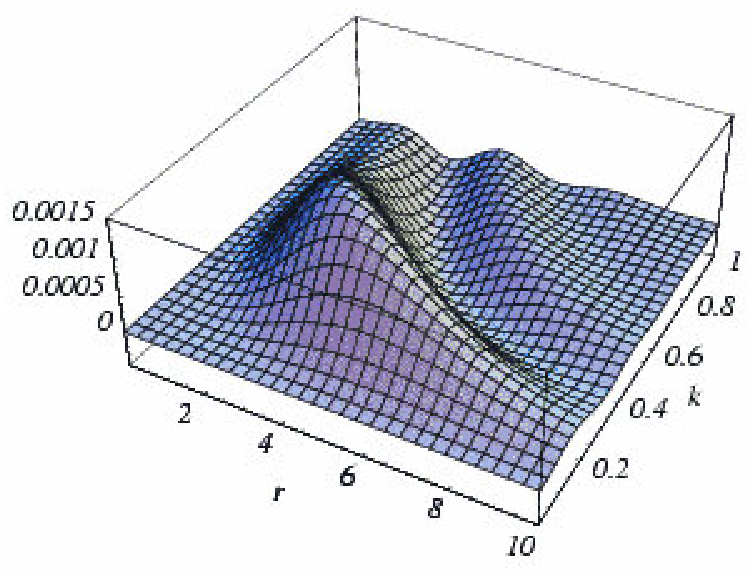}
\caption{\small The Wigner function of $2p_0$ state: Plot a) presents the
Wigner function for $\theta_1=\theta_2=0$;  b)
shows the same cross-section with the Wigner function is
multiplied by $r^2 k^2$. }\label{fig4a}
\end{figure}
\end{center}

The next example that we want to present in this paper is
the $\bf 2p_{0}$ state. In this case
\begin{equation}\label{d200n}
\mathbb{D}_{210}=
 \frac{2\exp( 2 i \vec{r}
\vec{k} )}{(2\pi a)^5} \op{b_1}\op{b_2}\, \left( \frac{\p^2}{\p
z^2}+4 i k_z \frac{\p}{\p z}\right)\,e^{-4i \vec{r}\vec{k}}\,.
\end{equation}
The form of this differential operator indicates that the
corresponding Wigner function will depend on the scalars
$(r,k)$ and two solid angles describing the orientations of
$\vec{r}$ and $\vec{k}$. The Wigner function of the $\bf 2p_{0}$
state is no
longer a function of scalars $\vec{r}^2$, $\vec{p}^2$ and
$\vec{r}\vec{k}$ is a consequence of the fact that
the wave function of this state
distinguishes the $z$ axis.
 As in all previous cases a close
form expression for the Wigner function
$W_{\psi_{210}}(\vec{r},\vec{k})$ can be calculated and
plotted. In Figure (\ref{fig4a}) we show cross-section of
 $W_{\psi_{210}}(r,k)$ for $\theta_1=\theta_2 =0$ and
the corresponding $r^2 k^2 W_{\psi_{210}}(r,k)  $ plot.
\vskip.2cm

\begin{center}
\begin{figure}[h]
\begin{picture}(0,0)(35,10)
\put(30,20){\makebox(0,0){a)}}
\put(195,20){\makebox(0,0){b)}}
\put(350,20){\makebox(0,0){c)}}
\end{picture}
\includegraphics[scale=.8]{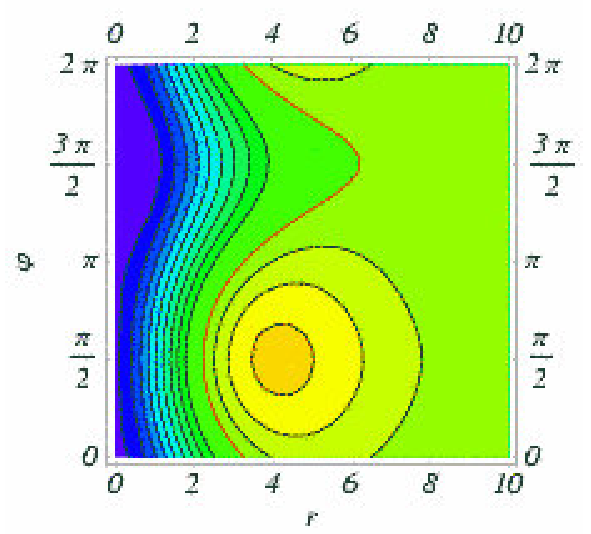}$\qquad$
\includegraphics[scale=.8]{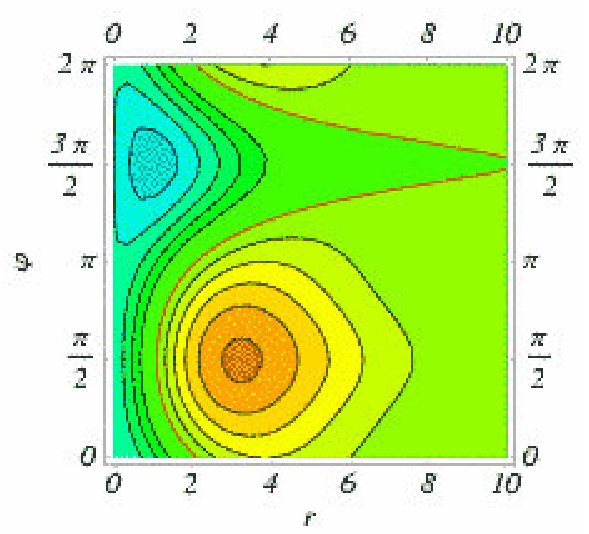}$\qquad$
\includegraphics[scale=.8]{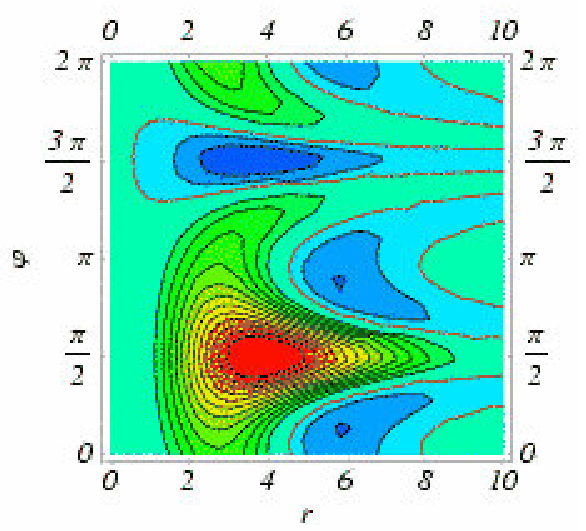}
\caption{\small The Wigner function of $2p_1$ state
 (we have chosen $\theta_1=\theta_2=\frac{\pi}{2}$, $\varphi_1=0$). Red line
 denotes a zero-level.
Plots a) and b) show cross-section of the Wigner function
for fixed values of momentum $k=0.1$ and $k=0.2$,
respectively. On the axes are $r$ and the angle between
$\vec{r}$ and $\vec{k}$. It is seen that maximum is
obtained when position and momentum vectors are
perpendicular, which agrees with classical intuition that
angular momentum has the maximum value for such geometry.
Plot c) presents similar cross-section of the Wigner
function multiplied by $r^2k^2$ for $k=0.5$. As we would
expect maximum is obtained for $\varphi=\frac{\pi}{2}$ and
$r\simeq 4$. }\label{fig4b}
\end{figure}
\end{center}

Finally, in the Figure (\ref{fig4b}) the Wigner function of 2p state with
 $m=1$ is shown.
We have plotted the cross-sections of
$W_{\psi_{211}}(\vec{r},\vec{k})$ for $\theta_1=\theta_2=\frac{\pi}{2}$,
$\varphi_1=0$
and  selected values of
$k$ as a function of $r$ and $\varphi_2$.  These plots show clearly that the
maxima of the Wigner
function   are reached for $\vec{k}$ perpendicular to
$\vec{r}$, which entirely agrees with classical intuition that
angular momentum has the maximum value for such geometry.
Thus, the semiclassical features begin to be
visible already in the 2p state. Of course the classical
features are much more pronounced for larger $n$.


In the remaining part of this Letter we illustrate the general
calculational scheme showing explicitly how one can calculate the
Wigner function for the ground state of the hydrogen atom. From
our derivation it will be clear that the method is general and can
be applied to all bound states of the hydrogen atoms. Although
most of the calculations for higher excited states look tedious,
simple symbolic differentiation of the fundamental HAI formula
leads to explicit expression for the the Wigner function with
arbitrary
quantum numbers.

We have found useful for our calculations to work with wave
functions in momentum representation \cite{hydrogen}. For
the wave function of the ground state we have:
\be
 {\mathbf{\tilde{\psi}}}_{100}(\vec{k})=
\frac{8\sqrt{\pi a^3}}{(1+k^2a^2)^2}
\label{moment}
\ee
It is a regular function that for large $k$ decreases as
$(ka)^{-4}$. Inserting this expression into the definition
of the Wigner function  in the momentum representation, Eq.
(\ref{wignerdef}), we obtain:
\begin{equation}
W_{\psi_{100}}(\vec{r},\vec{k})=\frac{2^9\pi a}{(2\pi a)^6}\int
d^3 q\,\frac{\exp(-2i\vec{r}(\vec{q}-
\vec{k}))}{[(\frac{1}{a^2}+q^2)(\frac{1}{a^2}+(\vec{q}-2\vec{k})^2)]^2}\,.
\label{Wigner momentum}
\end{equation}
The power of $q$ in the denominator can be reduced with the
help of differentiation over parameters. The following
expression for the Wigner function is obtained:
\begin{equation}
W_{\psi_{100}}(\vec{r},\vec{k})=\frac{2}{\pi^5
a^3}\frac{\partial^2} {\partial (b_1^2) \partial (b_2^2)}\int d^3 q\,
\frac{\exp(-2i(\vec{q}-\vec{k})\vec{r})} {(b_1^2+
q^2)(b_2^2+(\vec{q}-2\vec{k})^2)}\,, \label{Wignermom1}
\end{equation}
where $b_1$ and $b_2$ are running parameters to be fixed by
the Bohr radius at the end of all calculations. Now comes
the key element of the calculation. We recognize, that
apart of the phase factor, the structure of the integrand
in Eq. (\ref{Wignermom1}) has a remarkable formal
similarity to the product of two Klein-Gordon propagators
of quantum field theory in momentum space \cite{qft}. Due
to this analogy, we shall proceed with our calculations
using the standard propagator disentanglement techniques
introduced by Feynman and represented by the following
identity
\begin{equation}
\label{Fey} \frac{1}{AB} = \int_0^1du\,\frac{1}{\bigl[uA+(1-u)B
\bigr]^2}\,.
\end{equation}
with $A = b_1^2+q^2$ and $B = b_2^2+(\vec{q}-2\vec{k})^{2}$ to
rearrange  the integral in Eq. (\ref{Wignermom1}). The integral
becomes:
\ba W_{\psi_{100}}(\vec{r},\vec{k})&=&\frac{2}{\pi^5
a^3}\frac{\partial^2} {\partial (b_1^2) \partial (b_2^2)}\int_0^1 du\n\\
&\times&\int d^3 q\,
\frac{\exp\bigl[-2i(\vec{q}-\vec{k})\vec{r}\bigr]} {\bigl[u
(b_1^2+q^2)+(1-u)(b_2^2+(\vec{q}-2\vec{k})^2)\bigr]^2}
\label{WignerFeynman}\,.
 \ea
After rearrangement of terms and substitution:
$\vec{s}=\vec{q}-2(1-u)\vec{k}$ we get
\ba
 W_{\psi_{100}}(\vec{r},\vec{k})&=&\frac{2
e^{2i\vec{k}\vec{r}}}{\pi^5 a^3} \frac{\partial^2}
{\partial (b_1^2)
\partial (b_2^2)}\int_0^1 du  \n\\&& \times \int
d^3 s\, \frac{\exp\bigl[-2i(\vec{s}+2(1-u)\vec{k})\vec{r}\bigr]}
{[s^2+u\beta+(1-u)\gamma+4(1-u)u k^2]^2} \label{WignerFeynman2}\,.
\ea
We recognize in this expression the function $C(u)$
introduced in Eq. (\ref{Cdef}). This simplifies the
notation and the integral (\ref{WignerFeynman2}) then
becomes:
\begin{equation}
W_{\psi_{100}}(\vec{r},\vec{k})=\frac{2 e^{2i\vec{k}\vec{r}}
}{\pi^5 a^3 }\frac{\partial^2} {\partial (b_1^2)\partial (b_2^2)}\int_0^1
du \exp\bigl(-4i(1-u)\vec{k}\vec{r}\bigr)\int d^3 s\, \frac{\exp(-
2i\vec{s}\vec{r})}{\bigl[s^2+C(u)^2\bigr]^2}
\end{equation}
Fortunately, the integral over $d^3 s$ is elementary, we can use
the following formula:
\begin{equation}
\int d^3s\, \frac{\exp(-2i\vec{s}\vec{r})}{[s^2+C(u)^2]^2} =
\frac{\pi^2}{ C(u)} \exp\bigl(- 2 r C(u)\bigr).
\end{equation}
As a result of all these steps the only remaining integral is over
$u$. The final formula for the Wigner
function is thus given by:
\begin{eqnarray}
W_{\psi_{100}}(\vec{r},\vec{k})&=&\frac{2
e^{-2i\vec{k}\vec{r}}}{\pi^3 a^2} \frac{\partial^2}{\partial (b_1^2)
\partial (b_2^2)} \int_0^1 du \exp\bigl(i 4 u\vec{k}\vec{r}\bigr)
\frac{1}{C(u)} \exp\bigl(-2 r C(u)\bigr)\n \\
W_{\psi_{100}}(\vec{r},\vec{k})& =&
\mathbb{D}_{100}\bigg(\frac{\partial}{\partial
\vec{k}},\frac{\partial}{\partial b_1},\frac{\partial}{\partial
b_2} \bigg)\; I(r,k,\vec{r}\vec{k},b_1,b_2)\bigg{|}_{b_1=b_2=1/a}\,.
 \label{final}
\end{eqnarray}
as it has been advertised in Eq. (\ref{genformula}).

This method works for arbitrary state of the hydrogen atom.
The key concept in such calculations is to express a given
 state with quantum numbers $(nlm)$ in momentum
representation as a differential operator acting on the
ground state followed by a change of scale. It is worth
noting that these differential operators form an elegant
group theoretical structure, explained in e.g.
\cite{Barut}.
\vskip.2cm

In conclusion, we  have found and exploit formal similarity
between the eigenfunctions of nonrelativistic hydrogen atom
in the momentum representation and Klein-Gordon
propagators. This allowed us to find and discussed the
Wigner function for arbitrary bound state of hydrogen atom.
\\

\noindent
{\bf Acknowledgement}\\
L. $\!$P. thanks prof. Schleich for his encouragement and an interesting 
discussion.


\end{document}